\documentclass[a4paper]{jpconf}
\usepackage{times}
\usepackage{latexsym}
\usepackage{amssymb}
\usepackage{textcomp}
\usepackage{graphicx}
\usepackage{cite}
\usepackage{color}

\usepackage{fancyheadings}
\usepackage{lastpage}
\newcommand{\um}{\textmu m}

\begin{document}

\title{Microtomography on the ANATOMIX beamline at Synchrotron SOLEIL}

\author{T~Weitkamp$^1$, M~Scheel$^1$, J~Perrin$^1$, G~Daniel$^1$, A~King$^1$, V~Le~Roux$^1$, JL~Giorgetta$^1$, A~Carcy$^1$, F~Langlois$^1$, K~Desjardins$^1$, C~Menneglier$^1$, M~Cerato$^1$, C~Engblom$^1$, G~Cauchon$^1$, T~Moreno$^1$, C~Rivard$^{1,2}$, Y~Gohon$^3$, F~Polack$^1$}

\address{$^1$ Synchrotron SOLEIL, Gif-sur-Yvette, France}
\address{$^2$ TRANSFORM, INRAE, Nantes, France}
\address{$^3$ IJPB, INRAE, AgroParisTech, Versailles, France}

\ead{weitkamp@synchrotron-soleil.fr}

\begin{abstract}
The ANATOMIX beamline at Synchrotron SOLEIL, operational since 2018, is dedicated to hard X-ray full-field tomography techniques. Operating in a range of photon energies from approximately 5 to 50~keV, it offers both parallel-beam projection microtomography, in absorption and phase contrast, and nanotomography using a zone-plate transmission X-ray microscope. With these methods, the beamline covers a range of spatial resolution from 20~nm to 20~\um, expressed in terms of useful pixel size. The variable beam size of up to 40~mm allows users to image large objects. Here we describe the microtomography instrumentation of the beamline.
\end{abstract}

\section{Introduction}
The French synchrotron light source SOLEIL has recently built a long undulator beamline named ANATOMIX \emph{(Advanced Nanotomographic Imaging with Coherent X~Rays)} for X-ray microtomography and nanotomography. ANATOMIX completes the imaging methods available at SOLEIL by full-field tomography at lower photon energies and smaller length scales than PSICH{\'E} \cite{king2016rsi}, the other SOLEIL beamline offering full-field tomography as a main experimental method. The general layout and beam conditioning optics scheme have been reported elsewhere \cite{weitkamp2017jpcs}. Here we report on the main features of the experimental stations, in particular for microtomography, and on the current state of the instrument. The nanotomography station is described in more detail in a separate article in this volume \cite{scheel:elsewhere}.

In its two experiment hutches located, respectively, at distances of 170 and 200~m from the source, ANATOMIX can take the white X-ray beam or a monochromatic beam from a double-crystal Si-111 monochromator (DCM) or (once fully completed) from a double-multilayer monochromator. The nanotomography facility is exclusively operated with the DCM. For microtomography, the beam at sample position without any X-ray optics is approximately 20~mm wide and 15~mm high. It can be widened to more than 40 mm by a horizontally focusing double mirror located 35~m from the source. The first of the two mirror substrates is concave and its radius can be varied with a bending mechanism. When the mirror is not needed, it can be retracted from the beam. Beryllium refractive lenses in the optics hutches can be used to collimate the beam and increase the flux density at the sample position. Table~\ref{tbl:blpar} lists the main parameters of the beamline and its end stations.
\begin{table}[h]
  \caption{Key parameters of the ANATOMIX beamline.\label{tbl:blpar}}
  \small
  \begin{center}
    \begin{tabular}{ll}
      \br
      Experimental methods    & Microtomography (MT) in parallel-beam projection geometry \\
                        & Nanotomography with a transmission X-ray microscope (TXM)\\
                        & Local tomography available (MT, TXM)\\
                        & Extended field (=``half acquisition'', ``off-axis'') available (MT, TXM)\\[0.5ex]
Contrast modes          & Absorption contrast (MT, TXM)\\
                        & Inline phase contrast (MT); Zernike phase contrast (TXM) \\[0.5ex]
X-ray source            & In-vacuum, cryogenically cooled U18 undulator \\[0.5ex]
Spectral beam modes     & Filtered white beam \\
                        & Double-crystal monochromator Si-111 Bragg, vertical deflection \\
                        & Double-multilayer monochromator, vertical deflection$^{\mathrm{(a)}}$ \\[0.5ex]
Geometrical beam        & Double-bounce horizontally-reflecting mirror, removable, with bender \\
\quad{conditioning}     & Refractive lenses, removable \\[0.5ex]
Photon energies available
                        & \makebox[5.5em][l]{10 to 50 keV} (white-beam microtomography) \\
                        & \makebox[5.5em][l]{10 to 25 keV} (monochromatic microtomography) \\
                        & \makebox[5.5em][l]{7 to 17 keV}  (TXM nanotomography) \\[0.5ex]
Beam size at sample     & 40~$\times$~15~mm$^2$ (max.) \\
\quad (horizontal~$\times$~vertical) 
                        & 20~$\times$~15~mm$^2$ (without optics) \\[0.5ex]
Min.\ effective pixel size        & \makebox[3.5em][l]{20~nm}  (TXM nanotomography) \\
                        & \makebox[3.5em][l]{130~nm} (microtomography) \\[0.5ex]
Digital cameras         & \makebox[13em][l]{\it Make and model}\makebox[6em][c]{\it No.~of pixels}\makebox[4.5em][c]{\it Pixel size}\makebox[4em][c]{\it Speed}\\
                        & \makebox[13em][l]{Hamamatsu Orca Flash 4.0 V2}\makebox[6em][c]{2048$\times$2048}\makebox[4.5em][c]{6.5~\textmu m}\makebox[4em][c]{100~fps}\\
                        & \makebox[13em][l]{pco.dimax HS4}\makebox[6em][c]{2000$\times$2000}\makebox[4.5em][c]{11~\textmu m}\makebox[4em][c]{2277 fps}\\
                        & \makebox[13em][l]{pco.4000}\makebox[6em][c]{4008$\times$2672}\makebox[4.5em][c]{9~\textmu m}\makebox[4em][c]{5 fps}\\
                        & \makebox[13em][l]{Hamamatsu Orca Lightning$^{\mathrm{(b)}}$}\makebox[6em][c]{4608$\times$2592}\makebox[4.5em][c]{5.5~\textmu m}\makebox[4em][c]{30~fps}\\[0.5ex]
Detector optics         & Revolver-type optics (white or monochromatic beam): 5$\times$, 7.5$\times$, 10$\times$, 20$\times$ \\
\quad{and magnifications} & High-resolution optics (monochromatic beam only): 10$\times$, 20$\times$ \\
                        & Large-field optics (white or monochromatic beam): 0.48$\times$, 1$\times$, 2.1$\times$ \\[0.5ex]
Max.\ scan speed        & 1 tomography scan per second (standard rotation stage) \\
                        & 20 tomography scans per second (fast stage) \\
\br
\multicolumn{2}{l}{\footnotesize $^{\mathrm{(a)}}$Foreseen. $^{\mathrm{(b)}}$Under commissioning.}\\
\end{tabular}
\end{center}
\end{table}
\begin{figure}[htb]
  \centerline{\includegraphics[width=0.44\textwidth]{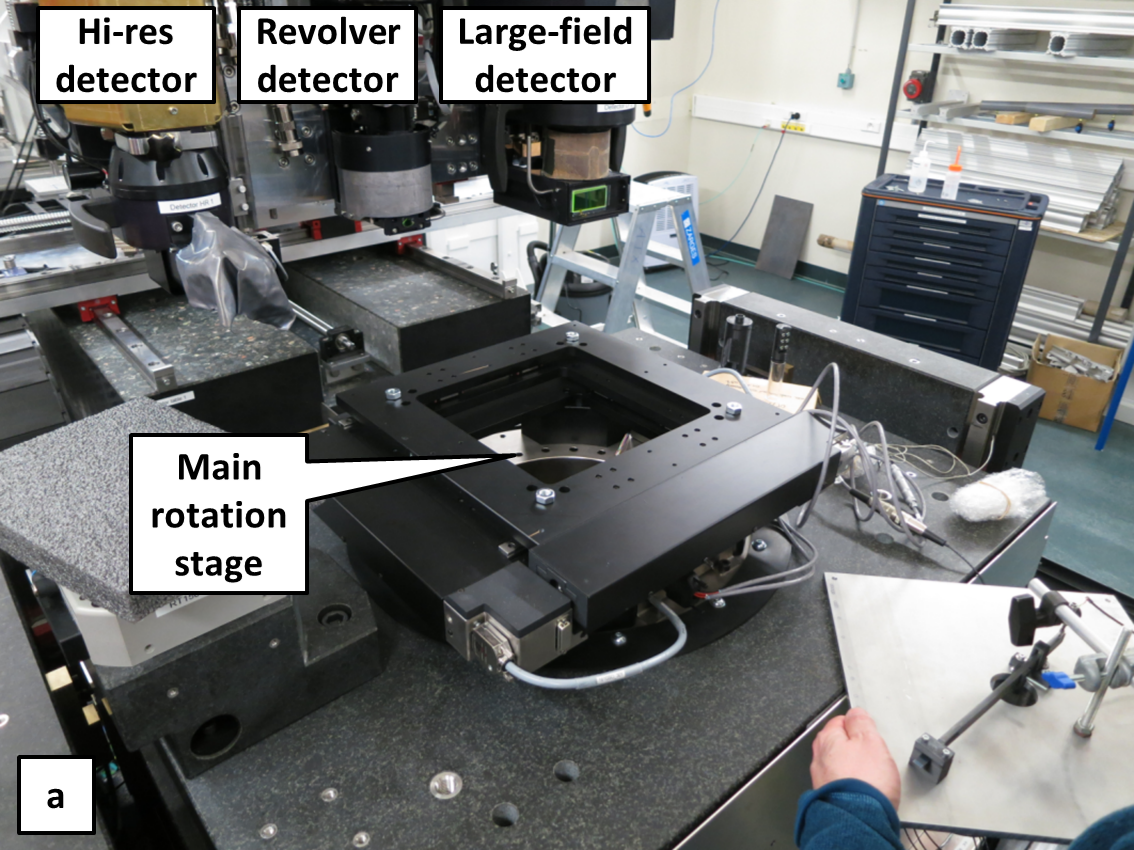}\hspace{0.02\textwidth}\includegraphics[width=0.44\textwidth]{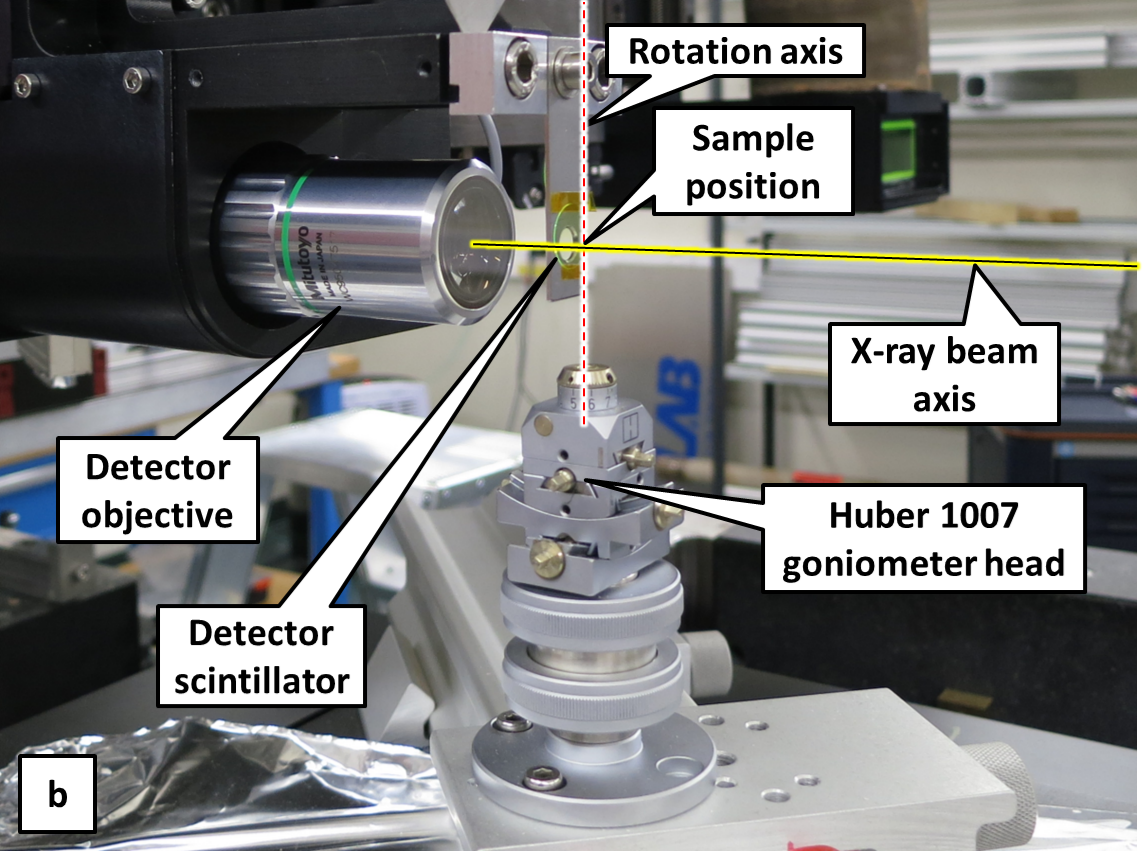}}
  \caption{(a) View of one of the two microtomography stations at ANATOMIX. In the center, the main rotation stage with XY stage (black frame) on the top; note the central hole of 250~mm usable diameter. In the background, three different detector optics are mounted. (b) Closeup of the high-resolution monochromatic detector and a standard mount for free-standing samples.\label{fig:microtomoone}}
\end{figure}
\begin{figure}[htb]
  \centerline{\includegraphics[width=0.9\textwidth]{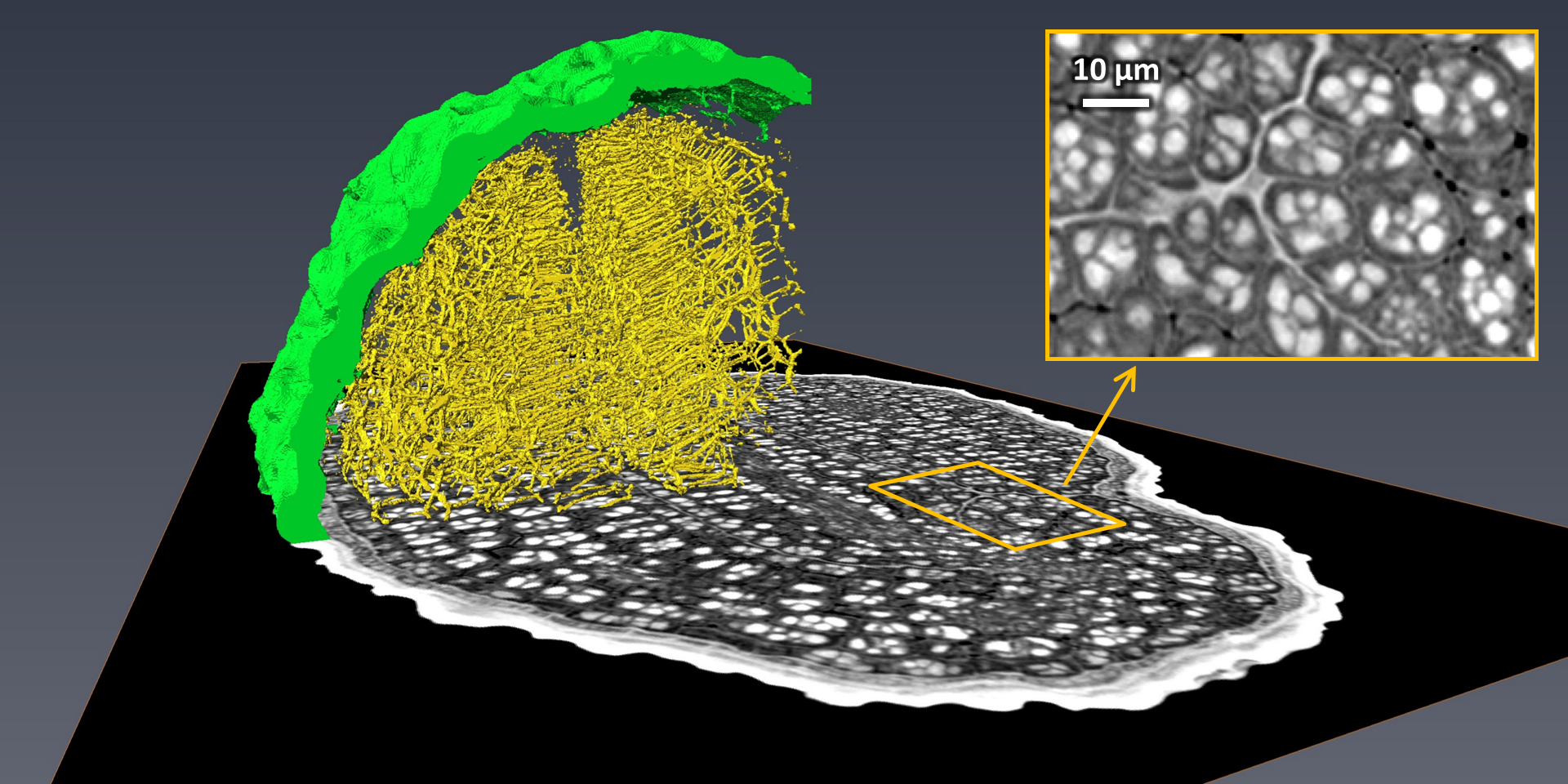}}
  \caption{\emph{Arabidopsis thaliana} seed (Columbia ecotype, col-0), data obtained by microtomography on ANATOMIX with a monochromatic beam (DCM Si-111) of 10~keV and a pixel size of 325~nm (2000 projections, total exposure time for this scan 200~s). The image shows a reconstructed longitudinal slice (gray) as well as part of the pore network (yellow) and the epidermis (green). The diameter of the grain is approximately 0.3~mm. In the detail inset on the upper right, protein storage vacuoles (white) can be discerned. This measurement was made in a coupled microtomography/nanotomography experiment, in the same session as the TXM results presented in Figure 3 of the paper by Scheel \emph{et al.} in this volume \cite{scheel:elsewhere}.\label{fig:thaliana}}
\end{figure}
\begin{figure}[htb]
  \centerline{\includegraphics[width=0.9\textwidth]{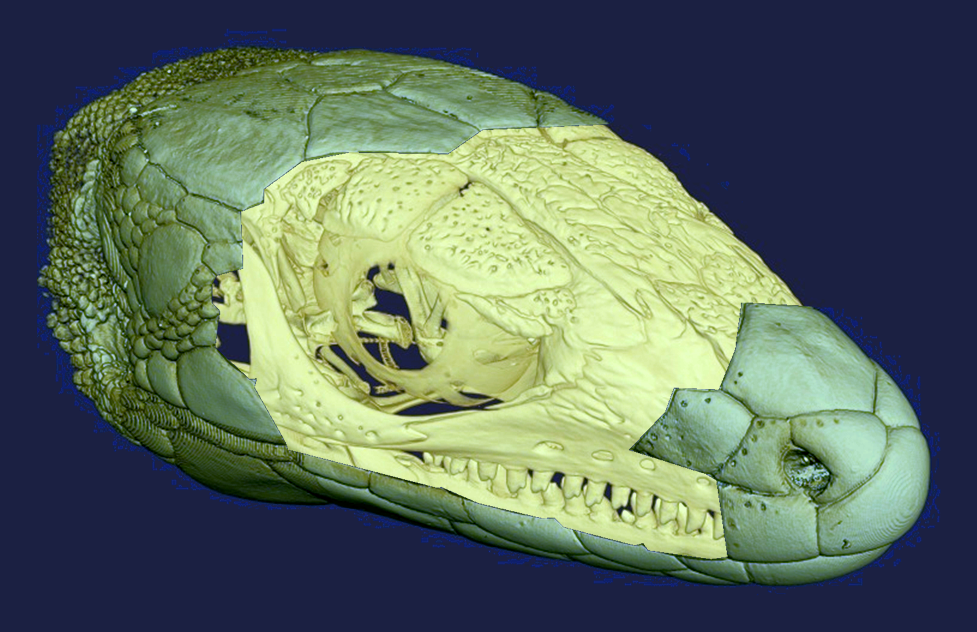}}
  \centerline{\includegraphics[width=0.9\textwidth]{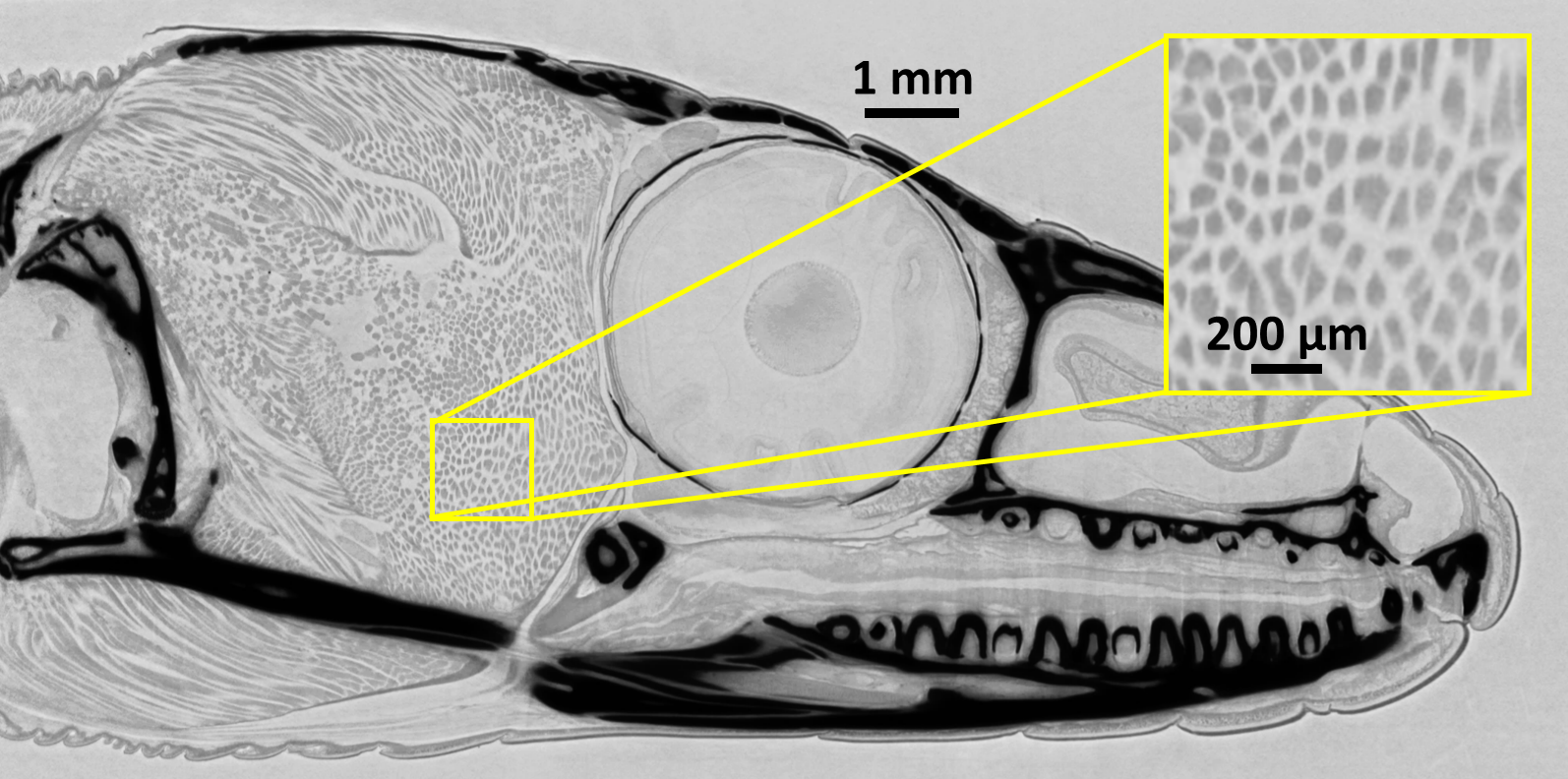}}
  \caption{Example of phase-contrast microtomography of a relatively large sample: a common wall lizard (\emph{Podarcis muralis,} dead animal immersed in ethanol). \emph{Top:} three-dimensional rendering of the head, recorded in a single tomography scan. \emph{Above:} Tomographic slice through the reconstructed volume. Data taken with a filtered white beam of around 25~keV central energy and a pixel size of 6.5~\um. The detector was placed 1.2~m behind the sample. Scan time: 7 minutes. A Paganin filter \cite{paganin2002} was applied for tomographic reconstruction. Note the muscle tissue visible in the detail inset on the right.\label{fig:lizard}}
\end{figure}

\section{Microtomography instrument layout and performance}

Two almost identical microtomographs are available, one in each of the two experimental hutches; they mainly differ by the available range of propagation distances between sample and detector, which is limited to 2.1~m in the first hutch, whereas in the second hutch distances up to 6~m can be realized. In principle it is also possible to place a sample on the microtomograph in hutch 1 and use a detector in hutch 2, which can increase the propagation distance up to 37~m. Hutch 1 also contains the TXM. While it is not possible to conduct two experiments simultaneously, the microtomograph in hutch 2 can be used to install and test complex setups while other experiments are running in hutch 1. Detector optics developed at SOLEIL \cite{desjardins:medsi2018} (see also Table~\ref{tbl:blpar}) give access to effective pixel sizes between 0.3 and about 20~\um\ in microtomography; these are complemented by the TXM with useful pixel sizes down to 20~nm.

Each of the microtomographs (Fig.~\ref{fig:microtomoone}) has two air-bearing tomography rotation stages. The standard stage is a model RT500S (LAB Motion Systems, Leuven, Belgium) with a central hole of diameter 250~mm for samples or environments requiring space far underneath the X-ray beam axis. An XY stage on top of the rotation allows for centering of the sample on the rotation axis. A smaller but faster rotation stage from the same supplier (model RT150S) is available for ultrafast tomography.

Both rotations are mounted on a common slab with a vertical travel range of 120~mm. Underneath, a horizontal transverse air-bearing translation driven by a linear motor can be used to change between rotation stages, center the rotation axis on beam and detector, and move the sample in and out of the beam for the acquisition of flatfield images.

While typical microtomography scans on ANATOMIX take a few minutes, scans can be much faster in experiments optimized for speed. Fast tomography can currently go up to approximately one scan per second with the RT500S standard stage and a pco.dimax HS4 camera (PCO AG, Kelheim, Germany). The RT150S stage is specified up to 13~rps, and first tests for ultrafast tomography have been conducted at acquisition rates of 18 tomography scans per second.

All scans are performed on the fly, i.e., the rotation stage does not stop rotating during acquisition; the synchronization of motorized axes and detector is usually obtained through the generic Flyscan architecture developed at SOLEIL \cite{leclercq:icalepcs2015}. The user interface for experiment control is based on Spyc, an iPython-based command-line and scripting environment at SOLEIL. Tomography-specific macros developed at the PSICH{\'E} beamline provide tools for easy sample alignment and scan setup, as well as a common user experience for both beamlines.

A local computer cluster for reconstruction and storage, with 5 computing nodes and 300~TB storage capacity reconstructs a standard (2048)$^3$-voxel volume in 3 minutes, using the reconstruction software PyHST2 \cite{mirone2014nimb} (ESRF, Grenoble, France). A Python script is used to pre-process data and set parameters of the reconstruction; here too, ANATOMIX benefits from developments on PSICH{\'E} and both beamlines provide a common user interface.

Figures \ref{fig:thaliana} and \ref{fig:lizard} show two examples of phase-contrast tomography, one at high resolution and one for a relatively large object, demonstrating some of the capabilities of the instrument. User experiments conducted on the beamline include studies from the vast domain of materials science, such as welding processes for fiber-reinforced composites \cite{mofakhami2020polymtest}, fundamental processes governing the drying of wood \cite{penvern2020prappl} or the way archeological textiles are preserved by mineralization in certain environments \cite{reynaud2020pnas}. Roughly 20\%\ of user projects involve sample environments for in-situ testing, for example for mechanical tests on novel materials for civil engineering \cite{ducoulombier2020strain} or to follow the evolution of the microstructure of frozen foods during freeze-thaw cycles \cite{zennoune2021foods}. The second big scientific domain served by the beamline is the biomedical field, which particularly benefits from the good phase-contrast conditions in studies of both soft tissue \cite{rodgers2021jneuroscimethods, rodgers2022jneuroscimethods, benzemzem2022ijms} and systems containing calcified tissue and prosthetic or other biocompatible materials \cite{jung2021ijms}.
  
\ack
ANATOMIX is an Equipment of Excellence (EQUIPEX) funded by the \emph{Investments for the Future} program of the French National Research Agency (ANR), project \emph{NanoimagesX}, grant no.\ ANR-11-EQPX-0031.

\section*{References}


\providecommand{\newblock}{}

\end{document}